\documentclass[pra,twocolumn,floatfix,superscriptaddress]{revtex4}

\usepackage{graphicx}
\usepackage{color}
\usepackage[normalem]{ulem}
\usepackage{braket}
\usepackage{amsmath}
\usepackage{verbatim}
\usepackage{amssymb}
\usepackage{hyperref}

\newcommand{\be}{\begin{equation}}
\newcommand{\ee}{\end{equation}}
\newcommand{\ba}{\begin{eqnarray}}
\newcommand{\ea}{\end{eqnarray}}
\newcommand{\nn}{\nonumber \\}

\def\p{\partial}

\def\exp{{\rm exp}}

\begin{document}

\title{Can quantum Monte Carlo simulate quantum annealing?}

\author{Evgeny Andriyash}
\affiliation{D-Wave Systems Inc., 3033 Beta Avenue, Burnaby BC
Canada V5G 4M9}

\author{Mohammad H.~Amin }
\affiliation{D-Wave Systems Inc., 3033 Beta Avenue, Burnaby BC
Canada V5G 4M9}
\affiliation{Department of Physics, Simon Fraser
University, Burnaby, BC, Canada V5A 1S6}

\begin{abstract}

Recent theoretical \cite{isakov2016understanding,Jiang:2017hee} and experimental \cite{denchev2016computational} studies have suggested that quantum Monte Carlo (QMC) simulation can behave similarly to quantum annealing (QA). The theoretical analysis was based on calculating transition rates between local minima, in the large spin limit using Wentzel-Kramers-Brillouin (WKB) approximation, for highly symmetric systems of ferromagnetically coupled qubits. The rate of transition was observed to scale the same in QMC and incoherent quantum tunneling, implying that there might be no quantum advantage of QA compared to QMC other than a prefactor. Quantum annealing is believed to provide quantum advantage through large scale superposition and entanglement and not just incoherent tunneling. Even for incoherent tunneling, the scaling similarity with QMC observed above does not hold in general. Here, we compare incoherent tunneling and QMC escape using perturbation theory, which has much wider validity than WKB approximation. We show that the two do not scale the same way when there are multiple homotopy-inequivalent paths for tunneling. We demonstrate through examples that frustration can generate an exponential number of tunneling paths, which under certain conditions can lead to an exponential advantage for incoherent tunneling over classical QMC escape. We provide analytical and numerical evidence for such an advantage and show that it holds beyond perturbation theory.

\end{abstract}

\maketitle

\section{Introduction}

Quantum annealing (QA) \cite{finnila1994quantum,kadowaki1998quantum,brooke1999quantum,Farhi2001,morita2008mathematical} is a computation scheme that harnesses quantum dynamics to find low-energy solutions of a problem. In QA, the system starts in a superposition of all logical states. Quantum fluctuations (i.e., superposition) are then reduced gradually, in a similar way as thermal fluctuations are reduced in thermal annealing, until a low-energy configuration is reached. In spin systems, this is commonly achieved by reducing the transverse field while the longitudinal terms in the Hamiltonian define the logical problem. A realistic quantum annealer interacts with a thermal environment, generating thermal transitions between the quantum eigenstates. As a result, during the annealing, the system initially follows equilibrium distribution up to some point. Beyond this point, it slowly deviates from equilibrium until its dynamics completely freeze \cite{Amin2015}. If it were possible to perform projective measurement in the middle of the annealing, the samples obtained would correspond to a Boltzmann distribution of the system's quantum Hamiltonian at the measurement point. At the end of the annealing, however, the solutions returned may not correspond to a Boltzmann distribution of the system Hamiltonian at the final or any intermediate point. Yet, they are expected to reflect the thermal nature of the quantum evolution.

Quantum Monte Carlo (QMC) simulations are classical algorithms designed to generate equilibrium statistics from a quantum Hamiltonian. In this paper, we only consider QMC algorithms with local updates such as path integral QMC with standard updates used for spin-glass simulations \cite{martovnak2002quantum}.  These algorithms can work as long as there is no {\em sign problem}, which requires the Hamiltonian be stoquastic \cite{bravyi2008complexity}, i.e., have no positive off-diagonal elements. If a QMC algorithm and a physical quantum system reach equilibrium for the same Hamiltonian, then the distributions that they generate will look the same, although their dynamics could be very different. It is therefore not possible to infer anything about the dynamics by just looking at the equilibrated probability distributions \cite{Amin2015}.

One can operate a QMC algorithm as an annealer by gradually reducing the transverse field in a similar way as in QA. The QMC algorithm would then reach equilibrium rather quickly at the beginning of the anneal, but towards the end equilibration becomes difficult, causing the algorithm to eventually deviate from equilibrium. This description looks similar to the one given above for QA, especially since the intermediate equilibrium statistics are the same. As a result, the two algorithms may sometimes show similar behavior even though the dynamics behind the equilibrium evolution may be different. This similarity has inspired some researchers to use QMC for predicting the behavior of QA \cite{santoro2002theory,heim2015quantum}. Recently, with commercial availability of the D-Wave quantum processing units (QPUs) \cite{Johnson2011}, it has become possible to compare the performance of a physical quantum annealer with QMC algorithms. Similarities in the behavior have been observed \cite{boixo2014evidence,denchev2016computational}, although differences have also been reported \cite{albash2015reexamining,albash2015reexamination}. Such similarities have raised the question of whether QMC is a viable simulation of QA. In other words, can QA provide any advantage beyond a prefactor in the scaling, such as the one observed in \cite{denchev2016computational}?

The physical resource behind QA is quantum tunneling. Coherent tunneling can support existence of eigenstates that are spread among many classical states via quantum superposition. Incoherent tunneling, on the other hand, allows random jumps (sometimes thermally assisted \cite{Dickson2013,Jiang:2017hee}) between localized states that are far away in Hamming distance. One can show that the rate of incoherent tunneling, $\Gamma_{\rm tunl}$, is proportional to the square of the multi-qubit tunneling amplitude, $g$, between the two localized states. More precisely \cite{Amin2008,neven2015computational},
 \be
 \Gamma_{\rm tunl} \propto {g^2 \over W}, \label{Gammatunl}
 \ee
where $W$ is the multi-qubit dephasing rate due to noise.

Quantum Monte Carlo dynamics, on the other hand, are based on Metropolis or other spin updates, which seem very different from the true quantum dynamics. In an interesting pair of papers, Isakov {\em et al.}~\cite{isakov2016understanding} and Jiang {\em et al.}~\cite{Jiang:2017hee} demonstrated that the calculation of QMC escape rate, $\Gamma_{\rm QMC}$, from one minimum to another has similarities to that for quantum tunneling amplitude. It was shown analytically that, in the large spin limit using WKB approximation,  for a fully-connected ferromagnet with uniform coupling, and when the temperature is low
 \be\label{QMCvsTunl}
 \Gamma_{\rm QMC} \propto g^2 \propto \Gamma_{\rm tunl},
 \ee
in leading exponential order. This means that the rate of transition between the local minima would scale with the number of qubits that are flipped (Hamming distance) in the same manner for both QMC and incoherent tunneling. Therefore, if this form of incoherent tunneling is the only quantum resource in QA, then QMC would behave similarly to QA not only in equilibrium statistics, but also in non-equilibrium dynamics. In other words, QMC would provide a viable simulation of QA. We emphasize that QMC can be considered a simulation of QA only if its performance is both statistically and dynamically similar to that of a quantum annealer. As such, we do not consider open boundary QMC \cite{isakov2016understanding} or diffusion Monte Carlo \cite{jarret2016adiabatic} as viable simulations of QA, because they cannot give correct equilibrium statistics by construction \footnote{When $T\ll 2 g$, open boundary QMC does simulate equilibrium statistics of zero temperature quantum Boltzmann distribution (ground state statistics). However, here we consider $T \gtrsim 2g$, which is a more realistic and interesting regime for incoherent tunneling.}. Such algorithms, however, can be viewed as classical optimization algorithms and benchmarked against other optimization methods, which may include QA.

As with any other quantum computation scheme, the power of QA must come from the ability to form large scale superposition and entangled states \cite{lanting2014entanglement} and not just random incoherent tunneling events. This is evident, for example, in the problems studied in Refs.~\cite{jarret2016adiabatic,somma2012quantum}, for which exponential advantage of QA compared to classical algorithms, including QMC, is established. Even for incoherent tunneling, the observation of Refs.~\cite{isakov2016understanding,Jiang:2017hee} only applies to the specific example considered. The authors of \cite{isakov2016understanding,Jiang:2017hee} were careful to mention a few situations where their argument could fail, but it was not clear if such situations can arise in real problems.

In this paper, we examine this question more thoroughly. We go beyond the large spin limit and WKB approximation, which have limited validity and only apply to special cases. We provide a much more general proof using perturbation theory. We show that (\ref{QMCvsTunl}) holds when there exists a single path for tunneling. In problems with many tunneling channels, however, quantum interference plays an important role in the tunneling process. A non-stoquastic Hamiltonian, for example, can produce destructive interference, which cannot be simulated by stochastic processes. In a stoquastic Hamiltonian, on the other hand, interference is always constructive and in principle can be represented by classical probabilities. However, as we shall show, reproduction of the interference effects by QMC requires overcoming topological obstructions. This is closely related to the topological obstructions for QMC discussed by Hastings and Freedman \cite{Hastings2013}. We show that, in the presence of constructive interference, quantum tunneling would escape with higher probability than QMC and the difference will increase with the number of homotopy-inequivalent tunneling paths. We introduce very simple examples in which tunneling can happen via such multiple paths and provide analytical and numerical results demonstrating the possibility of exponential superiority of incoherent tunneling over QMC escape within some limitations.

\section{Quantum tunneling}

\subsection{Problem setup}

In QA, one commonly considers a $N$-qubit Hamiltonian
 \ba
 H(s) &=& -A(s)\sum_{i=1}^N\sigma^x_i + B(s){\cal H}_P, \\
 {\cal H}_P &=& \sum_{i=1}^Nh_i\sigma^z_i + \sum_{i,j=1}^N
 J_{ij}\sigma^z_i\sigma^z_j, \label{HP}
 \ea
where $\sigma^{x,z}_i$ are Pauli matrices acting on qubit $i$, $s = t/t_a$, $t$ is time, $t_a$ is the annealing time, and $h_i$ and $J_{ij}$ are dimensionless parameters. The energy scales $A(s)$ and $B(s)$ are monotonic functions such that $A(0) \gg B(0) \approx 0$ and $B(1) \gg A(1) \approx 0$. Here, we only focus on incoherent tunneling and QMC escape at a particular point, $s^*$, instead of the full annealing process.  Incoherent tunneling happens when the tunneling amplitude is small ($s^*$ close to 1). In such a regime, one can use perturbation theory to approximate the tunneling amplitude as well as the QMC escape rate. We separate the Hamiltonian into unperturbed and perturbation parts:
 \be
 H=H_0 + V, \label{Hper}
 \ee
with
 \be
 H_0 = B(s^*) {\cal H}_P, \qquad
 V = - \Delta\sum_{i}\sigma^x_i,
 \ee
where the single qubit tunneling amplitude, $\Delta = A(s^*)$, is the small parameter in the expansion.

We consider a situation where the classical part of the Hamiltonian, $H_0$, forms a double-well potential with two minima, which we denote by ``up" ($\ket{\rm u}$) and ``down" ($\ket{\rm d}$) states, both at energy $E_0$. We assume that the two potential wells are identical and the energy gap between the two lowest energy levels is much smaller than their separation from other excited states $\delta E$. In this regime, the two lowest energy eigenstates are approximately $(\ket{\rm u}\pm \ket{\rm d})/\sqrt{2}$ and the energy gap between them is $2g$, where $g$ is the multi-qubit tunneling amplitude. We will also assume that the temperature $T$ is larger than the gap $2 g$ but much smaller than $\delta E$, so that only the two lowest energy levels are populated in thermal equilibrium. Our formalism can be extended to  cases when $T \gtrsim \delta E$ by taking into account thermally-assisted tunneling \cite{Jiang:2017hee}, but this is beyond the scope of this paper.

\subsection{Perturbative calculation of tunneling amplitude}

The perturbation Hamiltonian $V$ flips a single qubit in every application. If the Hamming distance between the two wells is $L$, then the lowest order perturbation that can generate off-diagonal terms between the minima is $L$. Define a path ${\cal P}_n=\{ {\bf s}^l\}_{l=0}^{n-1}$  as a sequence of $n$ states ${\bf s}^l=[s^l_1,s^l_2,\dots,s^l_N]$, $l=0,\dots,n{-}1$, in the computation basis,  with energies $E_l = H_0({\bf s}^l)$, where each pair of consecutive states differ by one bit-flip. The tunneling amplitude to the lowest order in $\Delta$ is given by \cite{amin2012approximate}
\ba
 g  =  \sum_{\rm {\cal P}_L}  \frac{\Delta^L}{ \prod_{l=1}^{L-1}(E_l-E_0)} . \label{D0}
\ea
where we have summed over all paths ${\cal P}_L$ that connect $\ket{\rm u}$ to $\ket{\rm d}$ through $L$ bit-flips with intermediate energies satisfying $E_l > E_0$.

\section{Quantum Monte Carlo}\label{subsec:QMCescape}

Quantum Monte Carlo simulation is based on the observation that equilibrium statistics of a $D$-dimensional (stoquastic) quantum Hamiltonian are equivalent to those of a ($D{+}1$)-dimensional classical Hamiltonian with appropriately chosen parameters (see Appendix \ref{Sec:QMC} for a brief introduction to QMC). The additional dimension, sometimes called imaginary time, has a periodic boundary condition. Therefore, configurations in  ($D{+}1$)-dimensional space can be viewed as  closed trajectories in $D$-dimensional space, called world-lines. QMC is a simulation of stochastic processes in the space of these world-lines satisfying the detailed balance condition. Equilibrium probabilities of the world-lines are proportional to their contributions to the partition function of the ($D{+}1$)-dimensional system.

In order to compute the sum of  equilibrium probabilities of the world-lines, we reduce the space of world-lines to the space of {\it loops}, as described in Appendix \ref{subsec:AppPertExpansion}. We define a loop ${\cal L}_n = \{{\bf s}^l \}$ as a {\em directed} closed path (as defined in the previous section) of length $n$ of states ${\bf s}^l$ with classical energies $E_l = H_0({\bf s}^l)$. The partition function can be expressed as a sum over the loops
 \be\label{eq:Z}
 Z = \sum_{n=0}^\infty \sum_{{\cal L}_n} e^{-F({\cal L}_n)},
 \ee
where $F({\cal L}_n)$ is the dimensionless free energy of the loop ${\cal L}_n$ (see Eq.~(\ref{FLoops})).

A new construction of the QMC algorithm directly in the loop space has recently been proposed in \cite{albash2017off}. Since the arguments that will follow are purely statistical, they naturally apply to this algorithm as well as other algorithms such as the Stochastic Series Expansion of Ref.~\cite{sandvik2003stochastic}.

\subsection{Boundary partition function}

Our goal is to find a relation between quantum Monte Carlo and quantum tunneling. A close connection between the two becomes evident if we expand the leading terms in $Z$ in powers of $g$. Let  $E_\pm=\tilde E_0 \pm g$ denote the two lowest  eigenvalues of the Hamiltonian with an energy gap $2g$ between them.  Here, $\tilde E_0$ is the renormalized lowest energy of each well (i.e., $E_0$ plus the self-energy corrections). Under the condition $2g{\ll} T{\ll} \delta E$, we can write
\be\label{eq:Zapprox}
Z \approx e^{-\beta E_+} {+} e^{-\beta E_-} \approx e^{-\beta \tilde E_0 } [ 2 +\beta^2 g^2 + O(\beta^4g^4)].
\ee
The first term is the sum of the contributions $e^{-\beta \tilde E_0}$ of each well, and the second term is the lowest order contribution of the tunneling amplitude $g$ to $Z$.

In order to understand the relation between (\ref{eq:Zapprox}) and QMC dynamics, we need to express  each term as a function of the loops. This can be achieved by regrouping the loops contributing to $Z$ by the number of times they travel between $\ket{\rm u}$ and $\ket{\rm d}$. We define  $R({\cal L})$ as the number of round trips that a loop ${\cal L}$ makes between the two minima. At low temperatures only the loops that pass through at least one of the minima will contribute to (\ref{eq:Z}) (see (\ref{FLoops})). We can therefore expand the partition function as
 \be\label{eq:ZR}
 Z = \sum_{r=0}^\infty \sum_{{\cal L}:\, R({\cal L})=r } e^{-F({\cal L})}.
 \ee
It turns out that terms with different $r$ in the above expansion correspond to different terms in  (\ref{eq:Zapprox}). The $r=0$ term contains loops that pass only through either $\ket{\rm u}$ or $\ket{\rm d}$ giving the local partition functions of each well. These loops act as self-energy terms renormalizing the energy $E_0$, hence
 \be \label{eq:Z0} 
 \sum_{{\cal L}:\, R({\cal L})=0 } e^{-F({\cal L})} = 2Z_0= 2e^{-\beta \tilde E_0 }.
 \ee
The term with $r=1$ is a sum over the loops that do a single round trip between the minima and give the $g^2$ contribution:
 \be\label{eq:Z1}
 \sum_{{\cal L}:\, R({\cal L})=1 } e^{-F({\cal L})} = Z_B = e^{-\beta \tilde E_0 } \beta^2 g^2. 
 \ee
These loops lie at the boundary between the two minima in the loop space. As such, we call $Z_B$ the {\em boundary partition function}.

To show that  (\ref{eq:Z0}) and  (\ref{eq:Z1}) hold, we use perturbation theory. To the lowest order perturbation,  $\tilde E_0=E_0$ and only loops of length $0$ contribute to $Z_0$ leading to $ Z_0 = e^{-\beta E_0}$, thus confirming (\ref{eq:Z0}). For $Z_B$, we sum over the loops that make a single round trip between the minima, which are of length $2 L$ to the lowest order perturbation.  Each loop ${\cal L}_{2L}$ can be split into two paths ${\cal P}_L$ and ${\cal P}'_L$, one from $\ket{\rm u}$ to $\ket{\rm d}$ and the other from $\ket{\rm d}$ to $\ket{\rm u}$. We therefore can write (see  (\ref{eq:probAtBarrier}))
 \ba\label{eq:Zb_pert}
 Z_B &\approx& \sum_{{\cal L}_{2 L} }  \frac{\beta^2  \Delta^{2L} e^{-\beta E_0} }{\prod_{l=1}^{L-1} (E_{l}{-}E_0) \prod_{l'=1}^{L-1} (E'_{l'}{-}E_0) }\nn
 &=& e^{-\beta E_0} \beta^2  \left( \sum_{{\cal P}_{L} }  \frac{\Delta^{L}}{\prod_{l=1}^{L-1} (E_{l}{-}E_0)} \right)^2 \nn
 &=& e^{-\beta E_0} \beta^2 g^2,
 \ea
where we have used (\ref{D0}). We show in Appendix \ref{subsec:QMCbeyond} that (\ref{eq:Z0}) and  (\ref{eq:Z1}) hold to all orders of perturbation theory.

Equations (\ref{eq:Z0}) and  (\ref{eq:Z1}) are the central results of this paper, which lie at the heart of relation (\ref{QMCvsTunl}) observed in Refs.~\cite{isakov2016understanding,Jiang:2017hee}. Their significance is that they connect the equilibrium population of certain loop configurations in QMC statistics to the tunneling amplitude:
\be\label{eq:gZB}
{Z_B \over Z_0} = \beta^2 g^2.
\ee
 This connection, however, is merely an equilibrium statistical property and does not directly translate into a QMC escape rate, which is a non-equilibrium process. We will show in the next subsection that during the escape process not all the loops participating in $Z_B$ will be visited according to the equilibrium statistics, due to topological obstructions, and hence (\ref{QMCvsTunl}) can be violated.

\subsection{Perturbative calculation of QMC escape}

We are interested in estimating the QMC escape rate for transitions between two local minima, $\ket{\rm u}$ and $\ket{\rm d}$, separated by a barrier. In principle, escape is a non-equilibrium process. However, if the escape rate is much smaller than the equilibration rate within the well that contains the initial local minimum, the local equilibrium statistics can determine the escape rate. This assumption is met for QMC in the regime $2g\ll T\ll \delta E$. We will apply this reasoning to obtain the leading contributions to the QMC escape rate. It is known that in the regime of intermediate-to-strong damping, the escape rate is dominated by the local equilibrium probability of saddle point (barrier) configurations in the energy landscape (see \cite{hanggi1990reaction} for a review). More precisely, the rate is proportional to the ratio of the total equilibrium probability of the world-lines in a small neighborhood of the barrier and that in the neghborhood of the local minimum. This is a multi-dimensional generalization of the celebrated Kramers' rate \cite{kramers1940brownian}. 

As the loop space is a reduction of the world-line space, we expect the escape rate $\Gamma_{\rm QMC}$ to be proportional to the ratio of the total equilibrium
probability of the loops ${\cal L}$ in a small neighborhood of the saddle point (barrier) $S$ and that of the local minimum $M$. The latter is the local partition function of the minimum $Z_0$, while the former  is commonly referred to as the {\it barrier partition function} $Z_{\rm barrier}$ in quantum state transition theory \cite{hanggi1990reaction}. The escape rate is, therefore, given by

\ba \label{Gm0}
 \Gamma_{\rm QMC} \propto  {\sum_{{\cal L} \in S} e^{-F({\cal L})} \over \sum_{{\cal L} \in M} e^{-F({\cal L})}} =  {Z_{\rm barrier} \over Z_0}.
\ea
We will use this relation to estimate QMC escape rate in different situations.

\begin{figure}[t]
\includegraphics[trim = 50mm 80mm 50mm 95mm, width=6cm]{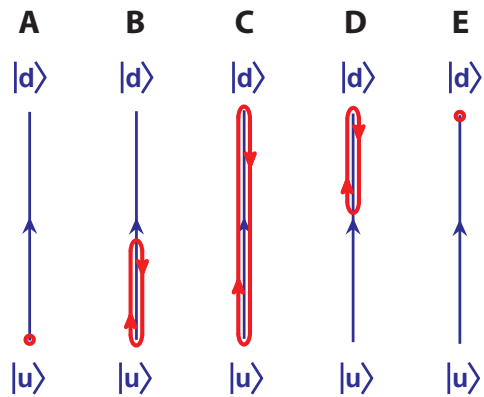}
\caption{Low temperature QMC escape process for a problem with one tunneling path. The blue lines represent the path connecting $\ket{\rm u}$ and $\ket{\rm d}$ in the computation basis. The QMC loop: starts from $\ket{\rm u}$ (A), stretches towards $\ket{\rm d}$ (B),  connects both minima (C), shrinks (D), and localizes in $\ket{\rm d}$ (E). The longest loop that happens in C sets the barrier height in Kramers' escape process.}
    \label{fig1}
\end{figure}

\subsubsection{Single tunneling path}

Let us first consider the case where there is only one dominant tunneling path connecting the two minima as depicted in Fig.~\ref{fig1}. The free energies of all the loops in Fig.~\ref{fig1}, except C, are given by (see Eq.~(\ref{eq:probNearBarrier})):
\ba
&& e^{-F({\cal L}_n)}\approx    \frac{\beta  \Delta^{n} e^{-\beta E_0} }{\prod_{l=1}^{n-1} (E_{l}-E_0)}.
 \ea
In this equation, $E_0$ is the lowest energy along the loop, which for the configurations in Fig.~\ref{fig1} coincides with the energy of the minima $\ket{u}$ and $\ket{d}$. Notice that temperature appears in $e^{-\beta E_0}$, while the length of the loop determines the order of perturbation $n$. Therefore, at low temperatures, the free energy cost of extending the loop, i.e., increasing $n$ without changing $E_0$, is lower than that of increasing its minimum energy. This means the escape will happen by extending the loop rather than by moving the zero length loop in Fig.~\ref{fig1} A upward along the tunneling path, which is equivalent to the classical thermal escape. As the loop extends from $\ket{\rm u}$ to $\ket{\rm d}$, its free energy  increases and reaches maximum at C, which determines the barrier. The sum over configurations in the neighborhood of C is precisely $Z_B$ and we have $Z_B = Z_{\rm barrier}$. Therefore, the QMC escape rate is given by 
\be\label{eq:GammaQMC}
\Gamma_{\rm QMC} \propto  {Z_B \over Z_0}  \approx  \beta^2  g^2,
\ee
confirming the relation (\ref{QMCvsTunl}), which was observed by \cite{isakov2016understanding}.

\subsubsection{Two tunneling paths}\label{subsubsec:twoPaths}

\begin{figure}[t]
   \includegraphics[trim = 45mm 100mm 45mm 100mm, width=7cm]{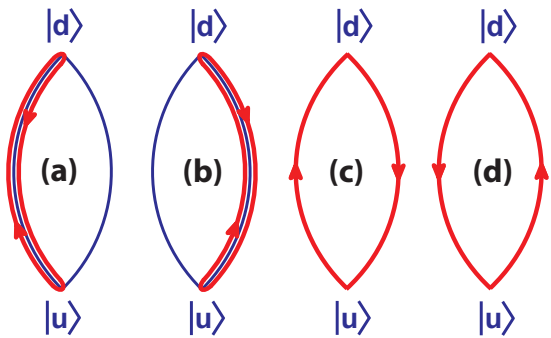}
\caption{A schematic diagram for two tunneling paths. The four configurations depict intermediate longest loops during QMC escape in the lowest order perturbation. The total equilibrium probability of these four loops is proportional to $g^2$.}
    \label{fig2}
\end{figure}

Let us now consider the case where there are two equivalent tunneling paths, as schematically depicted in Fig.~\ref{fig2} (see also Fig.~\ref{fig4} for an example). The two blue lines represent the tunneling paths. All other states outside the blue lines are assumed to have higher energies than those inside the paths and therefore do not participate in tunneling.  If $g_1$ represents the contribution of each path to the tunneling amplitude, we have $g = 2g_1$, and therefore
\be
\Gamma_{\rm tunl} \propto g^2 = 4g_1^2.
\ee

Figure \ref{fig2} also shows the loop configurations that make one round trip between the minima while staying on the low-energy tunneling paths. Because the loops are directional, (c) and (d) are distinct and there are four configurations that participate in $Z_B$. This means that $Z_B$ is given by the equilibrium population of the neighborhoods of these four configurations. However, as we shall see below only the left two configurations participate in QMC escape.

\begin{figure}[t]
   \includegraphics[trim = 45mm 30mm 45mm 40mm, width=6.5cm]{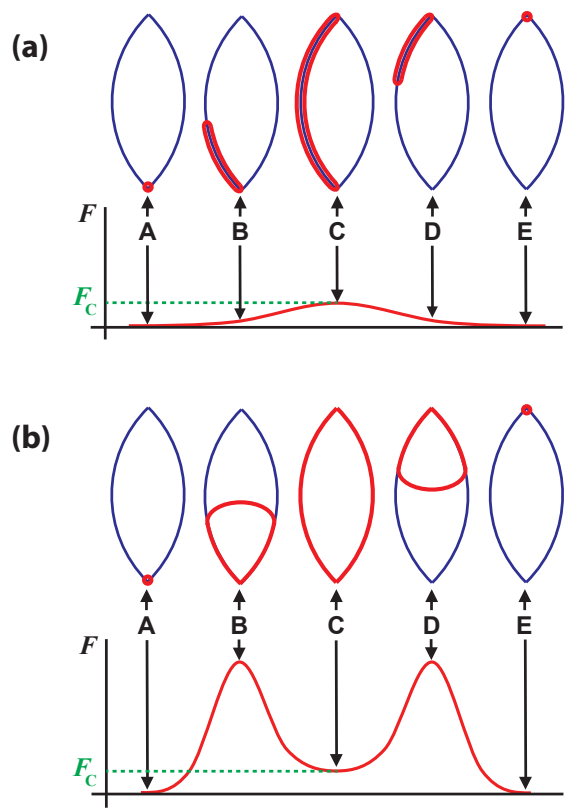}
\caption{QMC escape process along two tunneling paths when the initial loop stretches along (a) one path, and, (b) both paths. Schematic plots of the loop free energy $F$ during the escape is shown below the loop configurations. The free energy in the middle configuration (C) is the same for both panels (a) and (b). However, the maximum free energy (barrier) happens in configuration C in panel (a), but in configurations B and D in panel (b). As a result, the QMC escape rate in channel (b) is suppressed compared to (a).}
    \label{fig3}
\end{figure}

At low temperatures, there are two ways for the loop to stretch from $\ket{\rm u}$ to $\ket{\rm d}$ to facilitate escape:  stretching along a single path (Fig.~\ref{fig3}(a)), and stretching along both paths (Fig.~\ref{fig3}(b)). We will refer to these two ways as intra-path and inter-path escapes, respectively. The intra-path escape of Fig.~\ref{fig3}(a) is equivalent to the single path tunneling case considered above. The barrier in the free energy landscape corresponds to loop C and therefore the contribution of this channel to the QMC escape rate is proportional to $\beta^2 g_1^2$.

When the loop stretches along both paths, as in Fig.~\ref{fig3}(b), it cannot stay within the low-energy states (blue lines)  and has to pass through high-energy states during the escape process, as depicted in B and D. This means that the barrier in the free energy landscape can correspond to configurations B and D, instead of the middle (boundary) configuration C. If this barrier is much higher than $F_{\rm C}$, the contribution of this channel to the total escape rate will be significantly suppressed. Thus QMC escape rate will be dominated by the two intra-path tunneling channels of Fig.~\ref{fig2}(a) and (b), leading to
\be\label{QAdvtg}
\Gamma_{\rm QMC} \propto {1\over 2}{Z_B \over Z_0} ={1\over 2}\beta^2g^2 .
\ee
There is a factor of 2 quantum advantage for incoherent tunneling compared to QMC (\ref{eq:GammaQMC}). 

The above quantum advantage is closely related to the topological obstructions described by Hastings and Freedman \cite{Hastings2013}. In the subspace of the low-energy states (on the blue lines), the two tunneling paths are not homotopy-equivalent, i.e., they cannot be transformed into one another by deformation without leaving the subspace. As a consequence, the loops cannot stretch along both paths without leaving the subspace. 

We can also explain the above result by noticing that configurations C of  Fig.~\ref{fig2}(c) and (d) are not saddle points but local minima in the free energy landscape. Thus they don't contribute to $Z_{\rm barrier}$ but only to $Z_B$, leading to the relation $Z_{\rm barrier} = \frac{1}{2} Z_B$. In practice, there may be many saddle point configurations similar to B or D in Fig.~\ref{fig3}(b), whereas there is only one boundary configuration C. The inter-path tunneling is therefore suppressed only if (for lowest order perturbation)
\be\label{eq:freeEnergyComp}
\sum_{{\cal L}_n \in S} e^{-F({\cal L}_n)} \ll e^{-F_{\rm C}}.
\ee
In order for this relation to hold, the energy gap between the low-energy states and the excited states has to be large enough to offset the entropy difference between the two. As the system size increases, one has to increase the energy gap because the number of paths typically grows with the system size. When (\ref{eq:freeEnergyComp}) is violated, inter-path tunneling will not be suppressed anymore and the four configurations in Fig.~\ref{fig2} may determine the QMC escape rate, leading to (\ref{eq:GammaQMC}) with no factor of 2 advantage.

\subsubsection{Many tunneling paths}

The quantum advantage observed in (\ref{QAdvtg}) increases linearly with the number of homotopy-inequivalent tunneling paths ${\cal N}_{\rm paths}$ and therefore can lead to a scaling advantage. Because of constructive interference between the paths, quantum incoherent tunneling scales as  ${\cal N}_{\rm paths}^2$, whereas QMC escape scales as  ${\cal N}_{\rm paths}$, as long as the inter-path escape channels are forbidden. This leads to
\be\label{GammaNPaths}
\Gamma_{\rm QMC} \propto {\Gamma_{\rm tunl} \over {\cal N}_{\rm paths}} ,
\ee
which is different from (\ref{QMCvsTunl}) observed in \cite{isakov2016understanding,Jiang:2017hee}. As we shall see in the next section, in frustrated systems ${\cal N}_{\rm paths}$ can increase exponentially with the number of qubits that tunnel together. 

We would like to remark that although we have used perturbation theory, our conclusions hold beyond the lowest order perturbation expansion. In the next section, we provide numerical evidence that (\ref{GammaNPaths}) holds even when perturbation theory breaks down. Also, in Appendix \ref{subsec:QMCbeyond}, we provide a more general derivation of $Z_B$ beyond perturbation expansion. One may also generalize the above arguments to higher temperatures by taking into account thermally assisted tunneling events as in \cite{Jiang:2017hee}, but that is beyond the scope of this paper.

\section{Examples}

In this section, we introduce a few examples that capture the effects discussed in the previous section.

\subsection{Uniform ferromagnet}

Let us first consider the uniform fully-connected ferromagnet studied in \cite{isakov2016understanding,Jiang:2017hee}. The classical part of the Hamiltonian (\ref{HP}) is
 \be
 {\cal H}_P = -J\sum_{i,j=1}^N \sigma^z_i\sigma^z_j.
 \ee
The classical minima are therefore the two ferromagnetically oriented states, $\ket{\rm u}=\ket{\uparrow\uparrow\uparrow\dots\uparrow}$ and $\ket{\rm d}=\ket{\downarrow\downarrow\downarrow\dots\downarrow}$, with Hamming distance $L=N$, where $N$ is the number of qubits. Because of symmetry, there are $N!$ equivalent ways to flip the qubits from $\ket{\rm u}$ to $\ket{\rm d}$, and therefore $N!$ equivalent tunneling paths of length $N$. The subspace of the states covered by these paths includes all the $2^N$ logical states and one can deform any path into any other without leaving the subspace. In other words, all tunneling paths are homotopy equivalent, i.e., ${\cal N}_{\rm paths}=1$. The multiplicity of the paths therefore does not lead to topological obstructions and we obtain (\ref{QMCvsTunl}), in agreement with (although more general than) the result obtained in Ref.~\cite{isakov2016understanding}. This holds even for a nonuniform ferromagnet, as long as the loops ${\cal L}_{2N}$ that connect the two minima determine the barrier, i.e., the inter-path loops during the stochastic stretching do not create a bottleneck for the escape. Also, the above absence of  topological obstructions leading to (\ref{QMCvsTunl}) remains true at finite temperatures, as observed in \cite{Jiang:2017hee}.

\subsection{Frustrated ring}

\begin{figure}[t]
   \includegraphics[trim = 40mm 80mm 30mm 60mm, width=7cm]{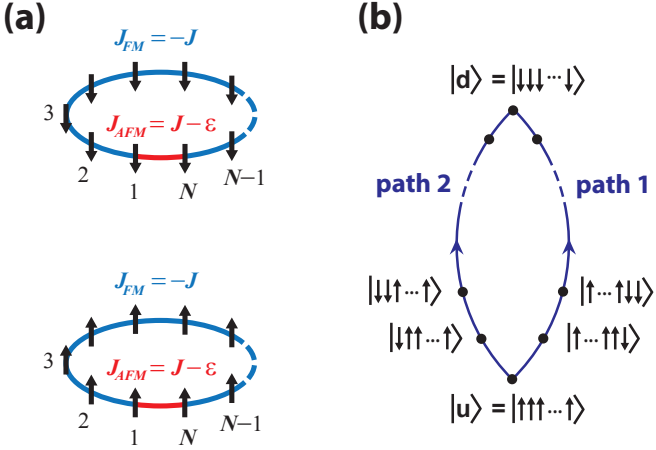}
\caption{(a) A frustrated spin ring. All couplings (blue lines) are ferromagnetic ($J_{FM}{=}-J$) except one (red line) that is antiferromagnetic ($J_{AFM}{=}J{-}\epsilon{<}J$).  The two lowest-energy classical states are shown: $\ket{\rm u}$ (bottom panel) and $\ket{\rm d}$ (top panel). (b) The two lowest-energy paths for tunneling from $\ket{\rm u}$ to $\ket{\rm d}$. In path 1 the qubits are flipped in the order $1,2,\dots,N$, while in path 2 the qubits are flipped in the reverse order. Other paths are energetically suppressed as long as $\epsilon \ll J$.}
    \label{fig4}
\end{figure}

As an example with two dominant tunneling paths, we consider a frustrated ring of qubits depicted in Fig.~\ref{fig4}(a). All couplers are ferromagnetic with $J_{FM}=-J$, except the one between qubit 1 and $N$ that is $J_{AFM}=J-\epsilon$. Since the antiferromagnetic coupling is weaker than the ferromagnetic ones, the classical ground states are states in which only this link is violated: $\ket{\rm u}=\ket{\uparrow\uparrow\uparrow \dots\uparrow}$ and $\ket{\rm d}=\ket{\downarrow\downarrow\downarrow\dots\downarrow}$. The Hamming distance between these two states is $L=N$. Starting from $\ket{\rm u}$, the energy cost of flipping one of the qubits on either sides of $J_{AFM}$ (qubit 1 or $N$) is $2\epsilon$. After that, the remaining qubits can be flipped sequentially, in the order $1{\to}N$ or $N{\to}1$ (see Fig.~\ref{fig3}(b)), with no energy cost until all qubits are flipped and state $\ket{\rm d}$ is reached. As a result, in the limit of $\epsilon {\ll} J$ and $\Delta {\ll} 4 J$, there are two dominant tunneling paths, with energy barrier $E_l{-}E_0=2\epsilon$. The tunneling amplitude in the perturbative regime ($\Delta<2\epsilon$) is given by
\ba
g = 2g_1 \approx 2 \frac{\Delta^N}{(2\epsilon)^{N-1}}.
\ea

All the states involved in these two tunneling paths have 1 violated coupling while all other states outside this subspace have at least 3 violated couplings. If the energy gap from the subspace to the outside ($\delta E=2(2J{-}\epsilon)$) is large compared to the tunneling barrier ($=2\epsilon$), which is the case when $J{\gg}\epsilon$, then the inter-path loops are energetically suppressed, leading to topological obstructions and quantum advantage (\ref{QAdvtg}).
Although $\delta E$ does not change with size, the entropy of the inter-path loops increases with $N$. At very large sizes the inter-path loops may find a chance to escape due to the large entropy and therefore reduce the factor of 1/2, unless $J/\epsilon$ is increased with $N$.

\begin{figure}[t]
   \includegraphics[trim = 50mm 110mm 50mm 90mm, width=7cm]{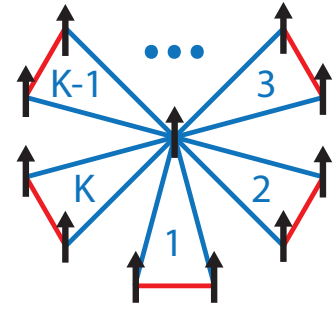}
\caption{An example problem with $K$ frustrated rings connected via a common node. Each blue line is a ferromagnetic coupler with $J_{FM}=-J$ and each red line is an antiferromagnetic coupler with $J_{AFM}=J-\epsilon$. The two degenerate classical ground states have ferromagnetic order with $2^K$ tunneling paths between them. Therefore, incoherent tunneling is up to $2^K$ times faster than QMC escape.}
    \label{fig5}
\end{figure}

\subsection{Shamrock}

To increase the number of tunneling paths, one can connect frustrated rings of the previous example so that they have to tunnel together. If $K$ frustrated rings are connected, since each allows two paths of tunneling, the total number of paths will be $2^K$. Assuming that all the inter-path QMC escape channels are topologically forbidden, we obtain
\ba
 \Gamma_{\rm QMC} \propto {\Gamma_{\rm tunl} \over 2^K}. \label{2Ksupremacy}
\ea
This is an exponential advantage of incoherent tunneling over QMC escape.

As an example, consider the graph depicted in Fig.~\ref{fig5}, which we call shamrock. Each frustrated ring has three qubits, one of which is shared among all the rings. Since the antiferromagnetic links (red lines) are the weakest, the classical ground states are: $\ket{\rm u}=\ket{\uparrow\uparrow\uparrow\dots\uparrow}$ and $\ket{\rm d}=\ket{\downarrow\downarrow\downarrow\dots\downarrow}$. A dominant path for tunneling from $\ket{\rm u}$ to $\ket{\rm d}$ is as follows: One outer qubit in each ring flips first. When half of the outer qubits (one per ring) are flipped, we reach the top of the energy barrier with the height equal to $2K\epsilon$. The central qubit can then flip with no energy cost. After that, the rest of the outer qubits will flip one by one. Since the order of flipping the outer qubits does not matter, there are $2^K$ possible ways to flip the qubits and therefore $2^K$ tunneling paths. The energy barrier for these paths is $2K\epsilon$, independent of $J$. The energies of all other excited states depend on $J$. Therefore, if $J{\gg}\epsilon$, the inter-path rings are energetically suppressed and we expect (\ref{2Ksupremacy}) to hold.

\begin{figure}[t]
   \includegraphics[trim = 50mm 90mm 50mm 90mm, width=7cm]{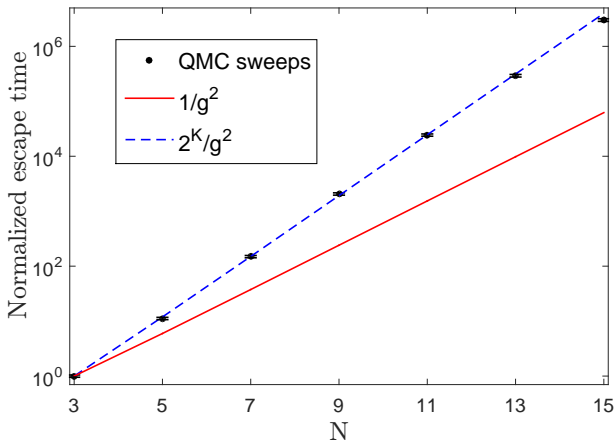}
\caption{Normalized escape time as a function of the number of qubits for the shamrock problems in Fig.~\ref{fig5}. Each curve is normalized to its value at $N=3$. Symbols represent the number of sweeps in QMC. The solid red line represents $1/g^2$, where $g$ is calculated via exact diagonalization. The dashed blue line plots $2^K/g^2$, where $K=(N{-}1)/2$ is the number of rings. The agreement between the symbols and the blue dashed line confirms (\ref{2Ksupremacy}). The simulation parameters are: $\Delta=A=0.5$, $B=1$, $\beta=20$, $J=6$, and $\epsilon=0.2$. Notice that the condition $\Delta<2\epsilon$ necessary for perturbation expansion is violated. The error bars represent statistical error for 1000 independent simulation runs.}
    \label{fig6}
\end{figure}

To test this numerically, we perform continuous-time QMC simulations starting from $\ket{\rm u}$ and probing when the escape happens. Similar to Ref.~\cite{isakov2016understanding}, we define escape time as mean first passage time, i.e. minimal time it takes QMC world-line to have some minimal (5\% in our calculations) support on state $\ket{\rm d}$. Figure \ref{fig6} shows the results of numerical calculations for shamrock graphs of Fig.~\ref{fig5} with $K=1$ to 7 rings ($N=3$ to 15 qubits). The parameters used in the simulation are given in the figure caption. The vertical axis in Fig.~\ref{fig6} shows the escape time normalized to its value for $K=1$. The symbols represent the number of QMC sweeps for each escape. The red line plots $1/g^2$ as a proxy for the incoherent tunneling time. The tunneling amplitude $g$ is obtained by calculating the minimum energy gap between the two lowest eigenstates using exact diagonalization of the Hamiltonian. Both QMC escape time and $1/g^2$ scale exponentially with $N$, but QMC scales worse. The blue dashed line plots $2^K/g^2$, which fits very well with the QMC sweeps confirming (\ref{2Ksupremacy}). Notice that with the parameters chosen, $\Delta > 2\epsilon$ and therefore perturbation theory does not hold while the exponential superiority predicted in  (\ref{2Ksupremacy}) remains valid.

Once again, one should be careful about the entropy of the inter-path loops as the shamrock graphs are scaled to larger sizes. Although the discrepancy between QMC escape and incoherent tunneling behavior still holds, to keep the above exponential advantage one may need to increase $J/\epsilon$ linearly with $N$.

\section{Conclusions}

We have used perturbation theory to compare incoherent tunneling and quantum Monte Carlo (QMC) escape rates in qubit systems. We have shown that the two can behave similarly when there is a single tunneling path. When multiple tunneling paths exist, constructive or destructive interference occurs. Constructive interference, which is the only one possible for stoquastic Hamiltonians, can lead to quantum advantage for incoherent tunneling in the presence of topological obstructions. The advantage can linearly increase with the number of tunneling paths. Frustration, which is a common feature in most hard problems, can produce multiple tunneling paths and therefore quantum advantage. We have shown through examples that in frustrated systems the number of tunneling paths can increase exponentially with the number of qubits that tunnel together. This can lead to an exponential quantum advantage as long as the topological obstructions remain effective, which requires careful examination of the entropy of the excited states involved in the topologically forbidden loops. We have also provided numerical evidence for such an advantage in the studied examples. 

While perturbation expansion was used in the main derivations of our work, the results hold beyond perturbation theory. We have demonstrated this in a numerical example and give a derivation in Appendix \ref{subsec:QMCbeyond} beyond perturbation theory.

A few remarks are in order. First, the rate given in (\ref{Gammatunl}) does not capture all the physics of incoherent tunneling. For example, resonant tunneling \cite{Amin2008,neven2015computational,lanting2014entanglement} and polaron effects \cite{neven2015computational}, which are important in the quantum tunneling analysis, are not captured by QMC, as demonstrated by, e.g., Albash {\em et al.}~\cite{albash2015reexamination}. Moreover, the $W$ in (\ref{Gammatunl}) is expected to scale as $\sqrt{N}$ if the noise is uncorrelated \cite{neven2015computational,lanting2010cotunneling}. There is also an additional factor of $N$ involved in QMC escape, because in every QMC sweep, $N$ qubits are flipped. Therefore, an additional $\sqrt{N}$ quantum advantage is expected beyond what is shown in Fig.~\ref{fig6}.

Finally, we should emphasize that incoherent tunneling is not the main quantum resource for QA. The ability to from large scale superposition and entanglement is the resource. Indeed, the exponential speedup of QA vs classical algorithms, demostrated by Somma {\em et al.} \cite{somma2012quantum}, would disappear if the algorithm relys merely on incoherent tunneling events.

\section*{Acknowledgment}

We acknowledge fruitful discussions with S.\ Boixo, J.\ Carrasquilla, F.\ Hamze, S.\ Isakov, H. Neven, V.\ Smelyanskiy, A.\ Smirnov, and M.\ Thom. We also thank F.\ Hanington, C.\ McGeoch, A. King, and S.\ Reinhardt for constructive comments on  our manuscript.

\appendix

\section{Quantum Monte Carlo simulation}\label{Sec:QMC}

In this appendix, we briefly review QMC algorithm and perturbation expansion of the partition function.

\subsection{Discrete-time QMC}

For Hamiltonian (\ref{Hper}), the partition function can be written as
 \ba
 Z = \text{Tr}\ e^{-\beta H} = \text{Tr} [e^{-\beta H\over M}]^M = \text{Tr}
 [e^{-\beta (H_0+V)\over M}]^M.
 \ea
Let $|{\bf s} \rangle \equiv |s_1s_2\dots s_N \rangle$ denote the classical
computational state with bit string $s_1s_2\dots s_N$ where $s_i=\pm 1$
for the $i$-th spin being up or down respectively. Inserting the
identity operator $I = \sum_{{\bf s}} |{\bf s}\rangle \langle {\bf s}|$ between each
power of $e^{-\beta H/M}$ we get
 \ba
 Z = \sum_{{\bf s}^1}\dots \sum_{{\bf s}^M} \langle {\bf s}^1|e^{-\beta (H_0+V)\over M}|{\bf s}^2\rangle
 \langle {\bf s}^2|e^{-\beta (H_0+V)\over M}|{\bf s}^3\rangle \nn
 \dots \langle {\bf s}^M|e^{-\beta (H_0+V)\over M}|{\bf s}^1\rangle . \hspace{2.7cm}
 \ea
The superscript in ${\bf s}^k$ denotes the $k$-th Trotter slice. In the large
$MT$ regime, one can approximately write $e^{-\beta (H_0+V)\over M}\approx
e^{-\beta V\over M}e^{-\beta H_0\over M}$. This is called Trotter approximation.
Since $H_0$ is diagonal in the computation basis, we have
 \ba
 \langle {\bf s}^k|e^{-\beta V\over M}e^{-\beta H_0\over M}|{\bf s}^{k+1}\rangle =
 \langle {\bf s}^k|e^{-\beta V\over M}|{\bf s}^{k+1}\rangle e^{-\beta H_0(k+1)\over M}. \nonumber
 \ea
On the other hand
 \ba
 \langle {\bf s}^k|e^{-\beta V\over M}|{\bf s}^{k+1}\rangle &=& \langle {\bf s}^k|\exp
 \left(\gamma \sum_{i=1}^N \sigma^x_i\right)|{\bf s}^{k+1}\rangle \nn
 &=& \prod_{i=1}^N \langle s_i^k|e^{\gamma \sigma^x_i}|s_i^{k+1}\rangle,
 \ea
where $\gamma = \beta \Delta /M$. Using $e^{\gamma \sigma^x_i} = \cosh
\gamma + \sigma_i^x \sinh \gamma$, we find
 \ba
 &&\langle \uparrow |e^{\gamma \sigma^x_i}|\uparrow \rangle
 = \langle \downarrow |e^{\gamma \sigma^x_i}|\downarrow\rangle
 = \cosh \gamma \nn
 &&\langle \uparrow|e^{\gamma \sigma^x_i}|\downarrow\rangle
 = \langle \downarrow|e^{\gamma \sigma^x_i}|\uparrow\rangle
 = \sinh \gamma.
 \ea
We can also write
 \ba
 && \cosh \gamma = \sqrt{{1\over 2} \sinh 2\gamma} \ e^{-{1\over 2} \ln (\tanh \gamma)} \nn
 && \sinh \gamma = \sqrt{{1\over 2} \sinh 2\gamma} \ e^{{1\over 2} \ln (\tanh \gamma)}.
 \ea
Substituting back
 \ba
 \langle {\bf s}^k|e^{-\beta V\over M}|{\bf s}^{k+1}\rangle = C^N
 e^{(\beta J^\bot/M)\sum_i s_i^ks_i^{k+1}}
 \ea
where
 \be
 J^\bot = -{M\over 2\beta} \ln \left(\tanh \gamma\right),
 \qquad C = \sqrt{{1\over 2} \sinh 2\gamma}
 \ee
Therefore
 \ba
 Z &=& C^{NM} \sum_{{\bf s}^1}\dots\sum_{{\bf s}^M} e^{-\beta \tilde{H}\over M}
 \ea
where
 \ba
 \tilde H = \sum_{k=1}^M \left[ H_0({\bf s}^k) - J^\bot \sum_{i=1}^N
 s_i^k s_i^{k+1} \right] \label{tildeH}
 \ea
is the $(D{+}1)$-dimensional classical Hamiltonian representing the
$D$-dimensional quantum system. Therefore, the equilibrium properties
of the two systems are the same.

\subsection{Continuous-time QMC}

To remove the Trotter error, one can take the limit $M\to \infty$. In practice it is enough to take $M \gg \beta \Delta N$. The partition function is given by the  sum over all closed trajectories of states (world-lines), each starting from ${\bf s}^1$ going through some intermediate configurations and ending in ${\bf s}^1$ again (periodic boundary condition):
 \ba
 Z &=& C^{NM} \sum_{\text{world-lines}} e^{-\beta \tilde{H}\over M}.
 \ea
The second term in (\ref{tildeH}) can also be written as
 \ba
\hspace{-0.5cm} -\sum_{i=1}^N \sum_{k=1}^M s_i^k s_i^{k+1}= 2\sum_{i=1}^N \sum_{k=1}^M \left({1- s_i^k s_i^{k+1} \over 2}\right) - NM. \label{sumNM}
 \ea
The last term can be absorbed into the coefficient:
 \ba
 C^{NM} e^{NMJ^\bot/M} = \left({1\over 2} \sinh 2\gamma\right)^{NM \over 2} e^{-{NM\over 2} \ln \left(\tanh \gamma\right)} \nn
 = (\cosh \gamma)^{NM}{\to}1, \quad \text{as } M{\to}\infty. \quad
 \ea
We also note that $(1 {-} s_i^k s_i^{k+1})/2$ is equal to 1 when $s_i^k \ne s_i^{k+1}$ and zero otherwise. Therefore the first term in (\ref{sumNM}) is $2n[\{{\bf s}\}]$, where $n[\{{\bf s}\}]$ in the number of spin-flips (solitons) along a the path $\{{\bf s}\}$. We can also replace $\tanh \gamma$ with $\gamma$ in this limit. The partition function can therefore be written in the form of path integral
 \ba \label{eq:pathInt}
 Z &=& \sum_{\text{world-lines}}   e^{\log(\gamma) n[{\bf s}]-\frac{\beta}{M}  \sum_{k=1}^M  H_0({\bf s}^k)}.
 \ea

\section{Perturbative expansion of partition function}\label{subsec:AppPertExpansion}

 We can organize the sum (\ref{eq:pathInt}) as a perturbative expansion:
 \ba\label{eq:pertExp}
 Z = \sum_{n=0}^{\infty} \left({\beta\Delta \over M}\right)^n \sum_{\text{world-lines}(n)}   e^{-\frac{\beta}{M}  \sum_{k=1}^M  H_0({\bf s}^k)} ,
 \ea
where the summation is done over configurations with $n$ spin flips in the imaginary time. Denoting the locations of those spin flips as $\mu_l, \ l=0,\dots,n-1, \mu_l \in [1, M]$ we note that spin orientation ${\bf s}^{\mu_l}$ between two consecutive spin flips $\mu_{l-1}$ and $\mu_l$ is constant, so as the classical energy $E_l=H_0[{\bf s}^{\mu_l}]$.  Defining $\lambda_l = (\mu_{l}{-}\mu_{l-1})/M$ in the limit $M \to \infty$ we can write
 \be
 \hspace{-0.1cm} \sum_{\text{world-lines}(n)}   e^{-\frac{\beta}{M}  \sum_{k=1}^M  H_0({\bf s}^k)}  =M^n   \sum_{ {\cal L}_n}\int_{\Sigma_n}  e^{-\sum_{l=0}^n \lambda_l E_l}  \nonumber
 \ee
where ${\cal L}_n=\{  {\bf s}^1,.., {\bf s}^n\}$ is a {\em directed} closed path in the computation basis with marked first state ${\bf s}^1$ and classical energies $E_l=H_0({\bf s}^l), E_0=E_1$ and $\Sigma_n \equiv \{\lambda_l \in {\rm I\!R}^{n+1}: \lambda_l \ge 0, \:\sum_{l=0}^n \lambda_l=1 \}$. Note that this expression can be obtained by directly expanding the partition function in powers of $\Delta$. We can sum up the contributions of all loops defined on a given set of states and differing only by the location of the marked state:
 \be
 \sum_{ \rm cyclic \: perm}\int_{\Sigma_n}  e^{-\sum_{l=0}^n \lambda_l E_l} = \int_{\Sigma_{n-1}}  e^{-\sum_{l=1}^n \lambda_l E_l} .
 \ee
When not all cyclic permutations correspond to unique ordered sets of states we have to divide by the order of the largest subgroup of the group of cyclic permutations that leaves the loop $ \{  {\bf s}^1, \dots,{\bf s}^n\}$ unchanged. This gives our final expression for the partition function as
\ba
 Z = \sum_{n=0}^\infty \sum_{{\cal L}_n}  \frac{(\beta\Delta)^n}{w({\cal L}_n)}  \int_{\Sigma_{n-1}}  e^{-\beta \sum_{l=1}^n \lambda_l E_l}.
 \ea
Contribution of each loop can be evaluated using Hermite-Genocchi formula  \cite{atkinson2008introduction}
\be
\int_{\Sigma_{n-1}}e^{-\beta\sum_l\lambda_l E_l}  = \beta^{-n+1} \sum_l \frac{e^{-\beta E_l}}{ \prod_{l' \ne l} (E_{l'}{-}E_l)}.
\ee
The integral is well-behaved even when $E_l=E_{l'}$. To express the result in the general case when some energies coincide we introduce multiplicities $m_l$ of the energies $E_l$, so that $E_l$ are unique energies along the path and  $H_0({\cal L}_n) = \{E_0, m_0; E_1, m_1;\dots\}$.
The partition function can now be written as
 \ba\label{ZLambda}
 Z = \sum_{n=0}^\infty \sum_{{\cal L}_n}  e^{-F({\cal L}_n)},
 \ea
where the free energy of the loop is given by
\be\label{FLoops}
e^{-F({\cal L}_n)}=  \frac{\beta\Delta^n}{w({\cal L}_n)}  \prod_{k} \frac{(-\p_{ E_k})^{m_k-1}}{(m_k-1)!}\sum_l  \frac{ e^{-\beta E_l}}{ \prod_{l' \ne l} (E_{l'}{-}E_l)}.
\ee
One can compute the contribution of a loop of length $n$ passing only once through the minimum of energy $E_0$:
\ba\label{eq:probNearBarrier}
&& e^{-F({\cal L}_n)}\approx    \frac{\beta  \Delta^{n} e^{-\beta E_0} }{\prod_{l=1}^{n-1} (E_{l}-E_0) },
 \ea
as well as  the contribution of  a loop connecting 2 minima,$\ket{\rm u}$ and $\ket{\rm d}$ via two paths, ${\cal P}_1(\ket{\rm u} {\to} \ket{\rm d})$ and ${\cal P}_2(\ket{\rm d} {\to} \ket{\rm u})$ of length $n_1$ and $n_2$
\be\label{eq:probAtBarrier}
 \hspace{-0.5cm} e^{-F({\cal L}_n)} \approx    \frac{\beta^2  \Delta^{n_1+n_2} e^{-\beta E_0} }{\prod_{l_1=1}^{n_1-1} (E_{l_1}^{(1)}-E_0) \prod_{l_2=1}^{n_2-1} (E_{l_2}^{(2)}-E_0)},
 \ee
where we have neglected corrections of order $T/\min(E_l^{(\cdot)}{-}E_0)$. Note that including those corrections will give
\be\label{eq:probAtBarrierCorr}
  \hspace{-0.5cm} e^{-F({\cal L}_n)} \approx    \langle \lambda_0 \rangle \frac{\beta^2  \Delta^{n_1+n_2} e^{-\beta E_0} }{\prod_{l_1=1}^{n_1-1} (E_{l_1}^{(1)}-E_0) \prod_{l_2=1}^{n_2-1} (E_{l_2}^{(2)}-E_0)},
 \ee
where $ \langle \lambda_0 \rangle = \frac{\int_{\Sigma_{n-1}}\lambda_0 e^{- \beta \sum_{l=0}^{n-1} \lambda_l E_l}}{\int_{\Sigma_{n-1}} e^{- \beta \sum_{l=0}^{n-1} \lambda_l E_l}}, 0 \le \langle \lambda_0 \rangle \le 1$ is the average time that the world-line spends in the state with energy $E_0$. To get (\ref{eq:probAtBarrier}) we assume that the temperature is low enough so that $\langle \lambda_0 \rangle \approx 1$.

\section{Proof of (\ref{eq:Z0}) and (\ref{eq:Z1})}\label{subsec:QMCbeyond}

In this section we show that the relations (\ref{eq:Z0}) and (\ref{eq:Z1}) hold to all orders of perturbaton theory. We use the approach of \cite{amin2012approximate} to compute the two lowest energy levels $E_\pm$. We separate  the Hilbert space into the low-energy subspace $\{\ket{\rm u},\ket{\rm d}\}$ and all other states. We define $P=\ket{\rm u}\bra{\rm u} + \ket{\rm d}\bra{\rm d}$ and $\bar P=I-P$ as projectors inside and outside of this subspace, where  $I$ is an identity operator. The effective low-energy Hamiltonian for the subspace $P$ is given by \cite{amin2012approximate}:
\be\label{eq:SchEq}
\left( E_0  + P V \bar P \frac{1}{E-\bar P \left( H_0 +  V \right)\bar P  } \bar P V  P \right) \psi = E \psi.
\ee
Assuming that the wells containing $\ket{\rm u}$ and $\ket{\rm d}$ are identical we can write this equation as
\be
\begin{bmatrix}  a(\delta E ) & b(\delta E ) \\  b(\delta E ) &  a(\delta E ) \end{bmatrix} \times \left[ \begin{array}{c} \psi_1 \\ \psi_2 \end{array} \right] = \delta E  \left[ \begin{array}{c} \psi_1 \\ \psi_2 \end{array} \right],
\ee
where $a(\delta E ) = \bra{u}  V \bar P \frac{1}{\delta E-(H_0-E_0) - \bar P V \bar P  } \bar P V  \ket{u}$ and $b(\delta E ) = \bra{u}  V \bar P \frac{1}{\delta E-(H_0-E_0) - \bar P V \bar P  } \bar P V  \ket{d}$.
Expanding $a$ and $b$ in perturbation $\bar P V \bar P$ gives the sum over all the paths of energy $E  > E_0$, connecting states $\ket{\rm u}{-}\ket{\rm u}$ and $\ket{\rm u}{-}\ket{\rm d}$. 

Using the free energy expression (\ref{FLoops})  and the above functions $a$ and $b$ we can write the local partition function $Z_0$ as
\be\label{eq:Z0_exp}
Z_0 = e^{-\beta E_0} + \sum_{m=1}^{\infty} \frac{(-\partial_{E_0})^{m-1}}{m!} \left( \beta e^{-\beta E_0}(-a_0)^m \right),
\ee
where we denote $a_0=a(0),b_0=b(0)$. Indeed, every loop that passes through the minimum (say $\ket{\rm u}$) $m$ times will contain $m$ segments that connect $\ket{\rm u}$ to itself through higher energy states. Since the contribution of each segment is given by $a_0$ and taking into account the cyclic permutation symmetry we obtain (\ref{eq:Z0_exp}).  Along the same lines one can express the boundary partition function as
\be\label{eq:ZB_exp}
Z_B =  \sum_{m=0}^{\infty} \frac{(-\partial_{E_0})^{m+1}}{m!} \left( \beta e^{-\beta E_0} b_0^2 (-a_0)^m \right),
\ee
because every loop that passes $(m+2)$ times through through the minima will have 2 segments that connect  $\ket{\rm u}$ and $\ket{\rm d}$ as well as $m$ segments that connect either $\ket{\rm u}$ or $\ket{\rm d}$ with itself. The contributions  of these segments to the loop free energy are given by $a_0$ and $b_0$ and since there are $(m+1)$ ways to distribute segments $a_0$ between the two minima we arrive at (\ref{eq:ZB_exp}). We can rearrange the expressions for $Z_0$ and $Z_B$ in a more suggestive way:
\ba\label{eq:Z_series}
&&Z_0 = e^{-\beta (E_0 + a_0)} {+}  \sum_{k=1}^{\infty}  \sum_{m=k+1}^{\infty} \frac{    e^{-\beta E_0}(-\beta)^{m-k}}{m k! (m{-}k{-}1)!}  \partial_{E_0}^{k} \left( a_0^m  \right)\nn
&&Z_B = \beta^2 e^{-\beta (E_0 + a_0)} b_0^2 +  \nn
&&+ e^{-\beta E_0}  \sum_{k=0}^{\infty}  \sum_{m=k}^{\infty} \frac{(m+1) (-\beta)^{m-k}}{k! (m-k)!}  \partial_{E_0}^{k+1}\left(  b_0^2 a_0^m \right).
\ea
One can already see that neglecting the derivative terms above we confirm (\ref{eq:Z0}) and (\ref{eq:Z1}) provided that $\tilde E_0 = E_0 + a_0$ and $g = b_0$. Since evaluating (\ref{eq:Z_series}) explicitely seems to be out of reach, we pursue a different approach. 

Note that (\ref{eq:ZR}) can be thought of as expansion of the full partition function in powers of $b_0$ and it's derivatives $b'_0,b''_0,..$. We compute the two lowest eigenvalues $E_\pm = E_0 + \delta E_\pm$ entering (\ref{eq:Zapprox}) in terms of functions $a$ and $b$ and show that the first two terms in (\ref{eq:Zapprox}) correspond to zeroth and second powers of $b$ respectively. The two lowest eigenvalues of the full Hamiltonian  are given by the lowest energy solutions to
\be\label{eq:eigenEq}
a(\delta E_\pm)   \pm b(\delta E_\pm) = \delta E_\pm.
\ee
Assuming that $b \ll a$ we can write the solution valid up to second order in $b$ as $E_\pm = \tilde E_0  \pm g $ with
\ba\label{eq:deltaE_g}
&& \tilde E_0 = E_0 +\delta E_0  + \frac{b b'}{(1-a')^2} + \frac{a'' b^2}{2 (1-a')^3} \nn
&& g = \frac{b}{1-a'},
\ea
where $a=a(\delta E_0),b=b(\delta E_0)$ and $\delta E_0$  is the smallest solution of $a(\delta E_0)   = \delta E_0$.
The full partition function can be expanded in powers of $b$ as
\ba
Z &\approx& e^{-\beta E_+} + e^{-\beta E_-} \approx\nn
&\approx& e^{-\beta  (E_0 +\delta E_0 ) } \left( 2 +\beta^2 g^2  + O(\beta b b')  + O(\beta a'' b^2)\right). \nonumber
\ea
At low temperature $\beta b b' \ll \beta^2 b^2$, $\beta a'' b^2 \ll \beta^2 a'b^2 <\beta^2 a'b^2$  and we obtain 
\ba
Z_0 &\approx& e^{-\beta  (E_0 +\delta E_0 ) } \nn
Z_B &\approx& e^{-\beta  (E_0 +\delta E_0 ) }\beta^2 g^2.
\ea

\bibliography{references}

\end{document}